\documentclass{jfm}

\usepackage{graphicx}
\usepackage{natbib}

\ifCUPmtlplainloaded \else
  \checkfont{eurm10}
  \iffontfound
    \IfFileExists{upmath.sty}
      {\typeout{^^JFound AMS Euler Roman fonts on the system,
                   using the 'upmath' package.^^J}%
       \usepackage{upmath}}
      {\typeout{^^JFound AMS Euler Roman fonts on the system, but you
                   dont seem to have the}%
       \typeout{'upmath' package installed. JFM.cls can take advantage
                 of these fonts,^^Jif you use 'upmath' package.^^J}%
      }
  \else
  \fi
\fi

\ifCUPmtlplainloaded \else
  \checkfont{msam10}
  \iffontfound
    \IfFileExists{amssymb.sty}
      {\typeout{^^JFound AMS Symbol fonts on the system, using the
                'amssymb' package.^^J}%
       \usepackage{amssymb}%
         \let\leq=\leqslant
         
      }{}
  \fi
\fi

\ifCUPmtlplainloaded \else
  \IfFileExists{amsbsy.sty}
    {\typeout{^^JFound the 'amsbsy' package on the system, using it.^^J}%
     \usepackage{amsbsy}}
    {\providecommand\boldsymbol[1]{\mbox{\boldmath $##1$}}}
\fi

\newsavebox{\astrutbox}
\sbox{\astrutbox}{\rule[-5pt]{0pt}{20pt}}

\newcommand{\VEC}[1]{{\bf #1}}
\newcommand{\VECADIM}[1]{\widetilde{\bf #1}}
\newcommand{\HAT}[1]{\widehat{\bf #1}}
\newcommand{\VECS}[1]{\boldsymbol{#1}}
\newcommand{\reff}[1]{(\ref{#1})}

\newcommand{\beq}{\begin{equation}}
\newcommand{\eeq}{\end{equation}}
\newcommand{\beqa}{\begin{eqnarray}}
\newcommand{\eeqa}{\end{eqnarray}}
\usepackage[usenames,dvipsnames]{color}

\title[Induced Diffusion in Bacterial Suspensions]{Induced Diffusion of Tracers in a Bacterial Suspension: Theory and Experiments}

\author[G. L. Mi\~no, J. Dunstan, A. Rousselet,  E. Clement and R. Soto]%
{G.\ns L.\ns M\ls I\ls \~N\ls O$^1$,\ns%
J.\ns D\ls U\ls N\ls S\ls T\ls A\ls N$^{2,3}$,\ns%
A.\ns R\ls O\ls U\ls S\ls S\ls E\ls L\ls E\ls T$^1$, \break
E.\ns C\ls L\ls \'E\ls M\ls E\ls N\ls T$^1$,\ns%
\and
 R.\ns S\ls O\ls T\ls O$^2$
  \thanks{Email address for correspondence: rsoto@dfi.uchile.cl}}
  
\affiliation{
$^1$PMMH-ESPCI, UMR 7636 CNRS-ESPCI-Universit\'e Paris 6 and Paris 7, 10 rue Vauquelin, 75005 Paris, France\\[\affilskip]
$^2$Departamento de F\'isica, FCFM, Universidad de Chile, Casilla 487-3, Santiago, Chile\\[\affilskip]
$^3$Department of Applied Mathematics and Theoretical Physics, University of Cambridge, Wilberforce Road, Cambridge CB3 0WA, UK}

\pubyear{2010}
\volume{650}
\pagerange{119--126}
\date{?; revised ?; accepted ?. - To be entered by editorial office}
\begin{document}

\maketitle

\begin{abstract}
The induced diffusion of tracers in a bacterial suspension is studied theoretically and experimentally at low bacterial concentrations. Considering the swimmer-tracer hydrodynamic interactions at low-Reynolds number and using a kinetic theory approach, it is shown that the induced diffusion coefficient is proportional to the swimmer concentration, their mean velocity and a coefficient $\beta$, as observed experimentally. The coefficient $\beta$ scales as the tracer-swimmer cross section times the mean square displacement produced by single scatterings. The displacements depend on the swimmer propulsion forces. Considering simple swimmer models (acting on the fluid as two monopoles or as a force dipole) it is shown that $\beta$ increases for decreasing swimming efficiencies. Close to solid surfaces the swimming efficiency degrades and, consequently, the induced diffusion increase. Experiments on W wild-type {\em Escherichia coli} in a Hele-Shaw cell under buoyant conditions are performed to measure the induced diffusion on tracers near surfaces. The modification of the suspension pH  vary the swimmers' velocity in a wide range allowing to extract the $\beta$ coefficient with precision. It is found that  the solid surfaces modify the induced diffusion: decreasing the confinement height of the cell, $\beta$ increases by a factor 4. The theoretical model reproduces this increase although there are quantitative differences, probably attributed to the simplicity of the swimmer models.
\end{abstract}


\section{Introduction}

An active suspension is a fluid containing autonomous swimmers such a bacteria, algae or artificial self-propelled entities.
In general, the presence of active swimmers changes the mechanical picture usually considered for passive suspension problems. When an organism swims, it interacts with the surrounding medium (i. e., the suspending fluid, other swimmers in the fluid and also the boundaries of the system). Consequently, balances of momentum and energy as well as the constitutive transport properties, are deeply modified by the momentum sources distributed in the bulk. When a microscopic cell swims in a fluid, usually viscous forces dominates the inertial terms, which is evidenced if we compare these two magnitudes by computing the Reynolds number ${\rm Re}=\frac{\rho V L}{\eta}$, where $V$ is a typical velocity, $L$ a typical length, and $\rho$ and $\eta$ are the density and the dynamic viscosity of the fluid, respectively. Swimmers, such as bacteria, with a typical length of $1\,\mu$m, and a propulsion velocity $V= 20\, \mu$m/s moving in water ($\rho = 1\,$g/cm$^3$ and $\eta = 10^{-2}\,$g/cm$\cdot$s), leads to a  Reynolds number in the order of $10^{-5}$ \citep{Purcell1977}. Under these conditions, autonomic propulsion is assured only if the time reversibility is broken during the motion \citep{Purcell1977,Golestanian2008}.

Since the swimming motion has no inertia, the total force and torque exerted by the swimmer on the fluid vanish.
Therefore, the leading long range hydrodynamic flow produced by a single swimmer should be governed at most by a dipole force. As a consequence, depending on the dipole polarity, many micro-organisms can be classified  under two categories: ``pushers'' and ``pullers'', depending if the dipole is tensile or contractile, respectively \citep{HernandezOrtiz2005, Saintillan2007, Baskaran2009}. From this point of view,  \emph{Escherichia coli} or \emph{Bacillus subtilis} can be classified as pushers and algae such as \emph{Chlamydomonas reinhardtii} as pullers.

\emph{Escherichia coli} (\emph{E. coli}) represents a good example of self-propelled swimmer. This widely studied bacterium will represent the active component in the solution studied in this work. \emph{E. coli} is a cylindrical cell with hemispherical endcaps, with $1\,\mu$m diameter and $2\,\mu$m length~\citep{BergPhysToday,Berg2004}. The motion is the consequence of the rotation of several helicoidal flagella, where each flagellum is linked to the cell membrane by a nanoscale motor. When all the motors turn in a counterclockwise (CCW) direction, the flagella rotates in a bundle and this pushes the cell steadily forward (what is called ``run''). When the motor rotation switches to clockwise (CW), the cell can change direction (known as a ``tumble'' process). The mean run interval is  typically 1 s, whereas the mean tumble time is around 0.1 s. The ratio between tumble and run depends upon the chemotactic signals~\citep{Berg2000}. Due to the drag anisotropy on the helicoidal flagella, its rotation produces a net force that pushes the fluid backward, while the head pushes the fluid in the opposite direction   \citep{Purcell1977,Lauga2009}. As consequence of the fluid incompressibility, the fluid is pulled toward the 
centre of the \emph{E. coli} body, creating two toroidal vortices around it.

The induced motion of the fluid produced by swimming microorganisms can be measured indirectly by observing the induced motion of tracers dispersed in the fluid.  The seminal work addressing the effect of active suspension on the enhancement of passive tracers  diffusivity, was presented by \citet{Wu2000}.
They conducted an interesting experiment: wild-type \emph{E. coli} as well as 4.5 and $10\,\mu$m diameter particles were trapped in a thin film \citep[see][figures 2 and 3]{Wu2000}.  When looking at the passive tracers, they found the presence of superdiffusion for $t<t_c$ and normal diffusion for $t>t_c$, where $t_c$ is a characteristic time representing the lifetime of coherent structures in the sample. They also studied the influence of the bacterial concentration in the diffusion of $10\,\mu$m diameter particles, showing the diffusion of these particles increases linearly with the bacterial concentration. They suggested that this effect could be related to the spontaneous formation of swirls in the bacterial bath.
In 2009, Leptos \emph{et al.}\nocite{Leptos2009} studied the enhancement of tracers diffusion into an aqueous medium, for a puller  type swimmer. They used \emph{Chlamydomonas reinhardtii} as the active component of the solution and $2\,\mu$m diameter beads as passive tracers. Their observation was performed far from the wall and they show a linear increase of the tracers diffusion with the volume fraction $\phi$. They found that the effective diffusivity $D_{\rm eff}$ for the passive tracers can be written as $D_{\rm eff}=D_B+\alpha \phi$, where $D_B$ is the Brownian diffusion value in the bulk (without swimmers) and, $\alpha$ can be defined (from dimensional analysis) as $\alpha = U^2 \tau = Ul$ (where $U$, $\tau$ and $l$ represent a characteristic advective velocity, encounter time, and advective length, respectively). In a recent work presented by \citet{Wilson2011},  Differential Dynamic Microscopy (DDM) was applied to characterised the bacterial motion in \emph{E. coli} suspensions. They show that the diffusion of non-motile bacteria is enhanced with the active fraction, given by the proportion of active swimmers, also establishing a linear relationship between these two quantities.

In 2011, we presented results obtained for passive particles near a solid surface in a bacterial suspension of density $n$ under non-buoyant conditions \citep{Mino2011}. We were able to distinguish two kind of bacterial motion: bacteria that exert a random motion and those  that move in ballistic way, called active swimmers. From this separation of populations sorted according to their motion, one can define a fraction of active swimmers $\phi_A$, and their mean velocity $V_A$. To characterise the activity at the surface we define the ``active flux'' $J_A=n_A V_A$, where $n_A=n  \phi_A$ is the active concentration.
Systematic experiments varying the bacteria concentration were done for two tracer sizes and in each case the mean diffusion of the passive tracer $D_P$  was obtained. Interestingly all data seem to group on a straight line
    \begin{equation}
    \label{eq:Diffusionatthewall}
    D_P=D_P^B+ \beta J_A,
    \end{equation}
where $D_P^B$ is the Brownian diffusivity close the wall in the absence of bacteria. The  parameter $\beta$ represents the slope that can be determined experimentally by a linear fit. A dimensional analysis of the equation (\ref{eq:Diffusionatthewall}) yields that this pre-factor is a length to the power 4. 

In the present work we present a kinetic theory analysis considering the swimmer-tracer hydrodynamic interactions, that explains the observed dependence of $D_P$ on $V_A$ and $n_A$, and provide an expression for $\beta$.
When a tracer is placed in the suspension, the swimmer-tracer scattering produce a net displacement that, as a consequence of the low-Reynolds number hydrodynamics as will be demonstrated in Sec. \ref{sec.theorybulk}, is independent on the swimmer velocity. In kinetic theory, the diffusion coefficient can be computed as the collision frequency times the net displacement squared. On the other hand, the collision frequency is proportional to the swimmer concentration, their mean speed and the cross section. Putting all together, results in an expression like \reff{eq:Diffusionatthewall}. The $\beta$ coefficient scales, then, as the cross section times the average displacement squared. As a result of the analysis, it turns out that an enhancement of the induced diffusion can be obtained near solid surfaces, calculations that are performed in Sec. \ref{sec.difusion.boundary}. Controlled experiments near solid surfaces, in which bacterial batches with different swimming speeds are prepared, are described in Sec. \ref{sec.experiments} showing the predicted enhancement of the diffusion near the solid surface and under confinement. The numerical comparison show good qualitative agreement. Finally, conclusions are presented in Sec. \ref{sec.conclusions}.

\section{Induced diffusion in the bulk} \label{sec.theorybulk}

Self-propelled objects agitate the fluid when they swim and, therefore, a passive object placed nearby will be transported by the fluid. At low Reynolds number, the transport is passive with no inertial delay. The experiments described in the Introduction and the new ones described in Sec. \ref{sec.experiments} show that, at low volume fraction, the induced diffusion is proportional to the swimmer concentration. This result implies that it should be possible to understand the induced diffusion as the additive effect of the interaction of the tracer with individual swimmers. Swimmer-swimmer interactions or the collective motion of the swimmers would lead to higher order contributions in the concentration to the induced diffusion. 
The hydrodynamic interaction of passive tracers with a single swimmer was studied in detail by \citet{Dunkel2010}. They showed that in the scattering process, the  tracer trajectories are loop-shaped, almost closed, with small total displacements. Here we consider the effect of these scattering processes in the induced diffusion of the tracers.
 
Two simple swimmer models will be considered in order to compute the induced diffusion on passive objects. At low Reynolds number, swimmers have no inertia and the net force and torque acting on them vanish. As a first approximation, a swimmer can be modelled as a force dipole. The second, more refined, model consists of two force monopoles separated by a finite distance, representing the head and tail of the swimmer. 

The Stokes equations allow to obtain the fluid velocity at a point $\VEC{r}$ produced by a force monopole $\VEC{F}$ located at $\VEC{r}'$ 
\beq
{u}_i(\VEC{r}) = J_{ij}(\VEC{r'}-\VEC{r}) F_j , \label{eq.Oseensolution}
\eeq
where $i$ and $j$ indicate the Cartesian coordinates, $J$ is the Ossen tensor
\beq
J_{ij}(\VEC{r}) = \frac{1}{8\pi\eta} \widehat{J}_{ij}(\VEC{r}) = \frac{1}{8\pi\eta} \frac{1}{ r} \left(\delta_{ij} + \frac{r_ir_j}{r^2}\right),
\eeq
$\delta_{ij}$ is the Kronecker symbol, $\eta$ is the fluid viscosity and the Einstein summation notation is used throughout the text \citep{HappelBrenner,KimKarilla}.
Thanks to the linearity of the Stokes equations, the velocity field produced by a force distribution is simple obtained by summing the different contributions to the fluid velocity.
Then, two monopoles located at $\VEC{r}_B\pm a\,\HAT{n}/2$ (that is, centred at $\VEC{r}_B$ and separated by $a$ in the direction $\HAT{n}$) produce a  velocity field
\beq
{u}_i(\VEC{r}) = \left[ J_{ij}(\VEC{r}_B+ a\,\HAT{n}/2 -\VEC{r}) - J_{ij}(\VEC{r}_B-a\,\HAT{n}/2 -\VEC{r} )\right] F_j , \label{vel.monopole}
\eeq
that in the dipolar limit reduces to
\beq
{u}_i(\VEC{r}) = J_{ij,k}(\VEC{r}_B-\VEC{r}) a F_j n_k , \label{vel.dipole}
\eeq
where $J_{ij,k}$ is the gradient of the Oseen tensor in the direction $k$
\beq
J_{ij,k}(\VEC{r}) = -\frac{1}{8\pi\eta} \left(\frac{\delta_{ij}r_k - \delta_{ik} r_j -\delta_{jk} r_i}{r^3} + 3 \frac{r_i r_j r_k}{r^5} \right).
\eeq

When a passive tracer is placed in the vicinity of a swimmer, it will move with a velocity equals to the fluid velocity at its centre, corrected by the Fax\'en contribution~\citep{HappelBrenner,KimKarilla}. This Fax\'en correction is proportional to the tracer radius squared and the Laplacian of the fluid velocity. In the experiments reported in \citet{Mino2011} the induced diffusivity does not present dependence on the tracer size for particle diameters of $1\mu$m and $2\mu$m, showing that in this regime the particle displacements are dominated by the fluid velocity at their center and the Fax\'en correction is subdominant. Accordingly, in what follows we will consider point-like tracers and the Fax\'en correction will be neglected. If precise calculations were needed, this correction should be added when the tracer and the swimmer are close, in which case the velocity field changes rapidly, but if this would be the case, also a more refined swimmer model would be necessary.  Then, the instantaneous tracer velocity is the fluid velocity at its position given either by \reff{vel.monopole} or \reff{vel.dipole}. 
The tracer motion can be made non-dimensional by the following choice of dimensions: lengths are scaled by the swimmer size $a$, $\widetilde{r}=r/a$, and dimensionless time $\widetilde{t}$ is defined in terms of the swimmer velocity $V_B$ by $\widetilde{t}=t a/V_B$. The resulting tracer equation of motion is
\beq
\frac{d \widetilde{r}_i}{d\widetilde{t}} = \gamma^{-1} G_i(\VECADIM{r}_B-\VECADIM{r}), \label{ec.motion}
\eeq
where the dimensionless parameter
\beq
\gamma=8 \pi V_B a \eta/F \label{eq.gamma}
\eeq
is a measure of the swimming efficiency and
\beqa
G^{\rm monopoles}_i (\VEC{x}) &=& \left[\widehat{J}_{ij}(\VEC{x}+\HAT{n}/2)- \widehat{J}_{ij}(\VEC{x}-\HAT{n}/2)\right] n_j \label{Gmonopoles}\\
G^{\rm dipole}_i (\VEC{x}) &=& \widehat{J}_{ij,k}(\VEC{x})  n_j n_k \label{Gdipoles}.
\eeqa
We have assumed, as it is usually the case, that the direction of the force is parallel to $\HAT{n}$, with $\VEC{F}=F\HAT{n}$. Therefore, $F$ and $\gamma$ are positive (negative) for pushers (pullers).

It is worth noticing that the swimming efficiency $\gamma$ depends on the geometry of the swimmer but it does not depend directly on the fluid viscosity or the activity of the swimmer (e.g. the flagella rotation speed for {\em E. coli}). 
Indeed, if the swimmer increases its activity, by the linearity of the Stokes equations, it will swim more rapidly keeping $\gamma$ unchanged. Also, if the activity is fixed, and the swimmer is placed in a more viscous fluid, its swimming velocity will decrease and $\gamma$ is not modified.
In the case of {\em E. coli} in water, using the typical velocity of $V_b=20\,\mu{\rm m/s}$ \citep{Berg2004}, the measured dipole distance $a=2\,\mu{\rm m}$ \citep{Drescher}, and a fitted value $F=1.13\, {\rm pN}$, calculated by \cite{Dunstan2012} using a two-sphere model, it gives an efficiency of $\gamma_{E. coli}=0.89$. The swimming efficiency changes, however, close to fixed solid surfaces where, for a fixed force, the swimmer moves slower than in the bulk. This issue will be exploited in Sect. \ref{sec.difusion.boundary} as it will be shown to increase the induced diffusion. 
There is, finally, the possibility that dynamically the swimmer changes its efficiency when moving in a more viscous fluid. Indeed, it has been reported that increasing the viscosity, some bacteria swim faster, effect that is explained by a modification of the flagella geometry due to a balance of elastic and viscous stresses \citep{Schneider74,Atsumi96,Chattopadhyay06}. To simplify the analysis and because we have no evidence that there are conformational changes in our experiments, we will assume that the geometry remains fixed.

\begin{figure}
\begin{center}\includegraphics[width=.9\columnwidth]{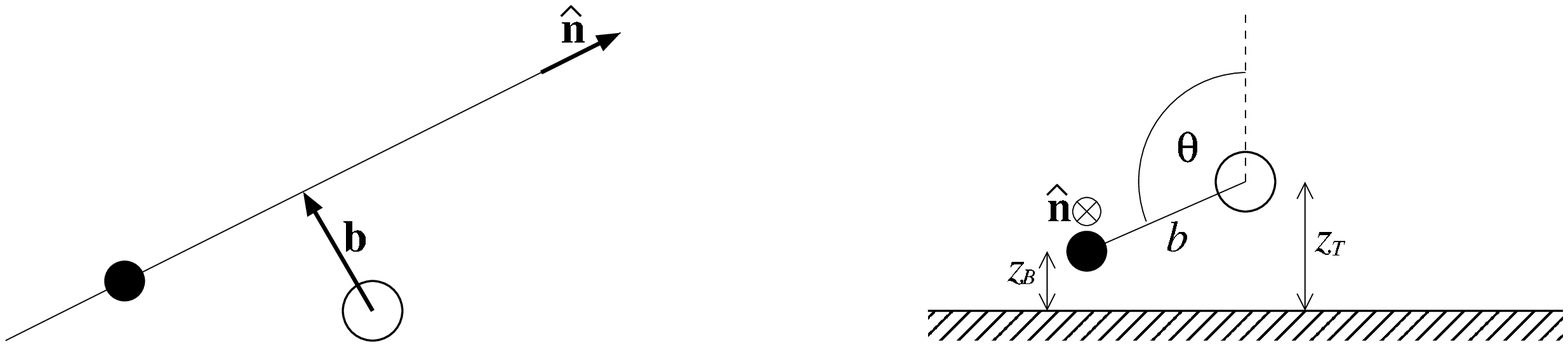}\end{center}
\caption{Parameters describing the scattering geometry between a swimmer (solid circle) and a tracer (open circle) in the bulk (Left) and close to a solid surface (Right). The swimmer moves in the $\HAT{n}$ direction (parallel to the surface when close to it). The vector $\VEC{b}$ points from the  initial position of the tracer to the position of minimal approach of the straight trajectory and its magnitude is $b$, the impact parameter. Close to a solid surface, the tracer initial height is $z_T$, the angle $\theta$ gives the horizontal displacement of the swimmer trajectory with respect to the initial position of the tracer and the height of the swimmer is $z_B$. }
\label{fig.paramsscatt}
\end{figure}

To integrate Eq. \reff{ec.motion} the swimmer motion must be specified. As the dipolar interaction decays at large distances, the swimmer motion can be assumed to be in a straight line, with constant speed, during all the scattering process. That is, we neglect the circular motion bacteria perform near surfaces and, consistently, the torque dipole created by the rotation of the head and flagella are neglected as well. Finally, we  neglect tumbling during the scattering process. As usual when describing scattering processes, the swimmer trajectory is parametrized by the direction of motion $\HAT{n}$, the  dimensionless impact parameter $\widetilde{b}=b/a$ and the azimuthal angle $\theta$, that define completely the vector $\VECADIM{b}$ that goes from the initial position of the tracer to the point of minimal approach of the swimmer trajectory (see Fig. \ref{fig.paramsscatt}). Then, the full swimmer trajectory in dimensionless units is parametrized by $\VECADIM{r}_B =
\widetilde{t}\HAT{n}+\VECADIM{b}$, with $-\infty<\widetilde{t}<\infty$, and the origin has been placed at the initial position of the tracer. Then, the tracer equation of motion reduces to
\beq
\frac{d \widetilde{r}_i}{d\widetilde{t}} = \gamma^{-1} G_i(t\HAT{n}+\VECADIM{b}- \VECADIM{x}). \label{ec.motion2}
\eeq

\begin{figure}
\begin{center}\includegraphics[width=\columnwidth]{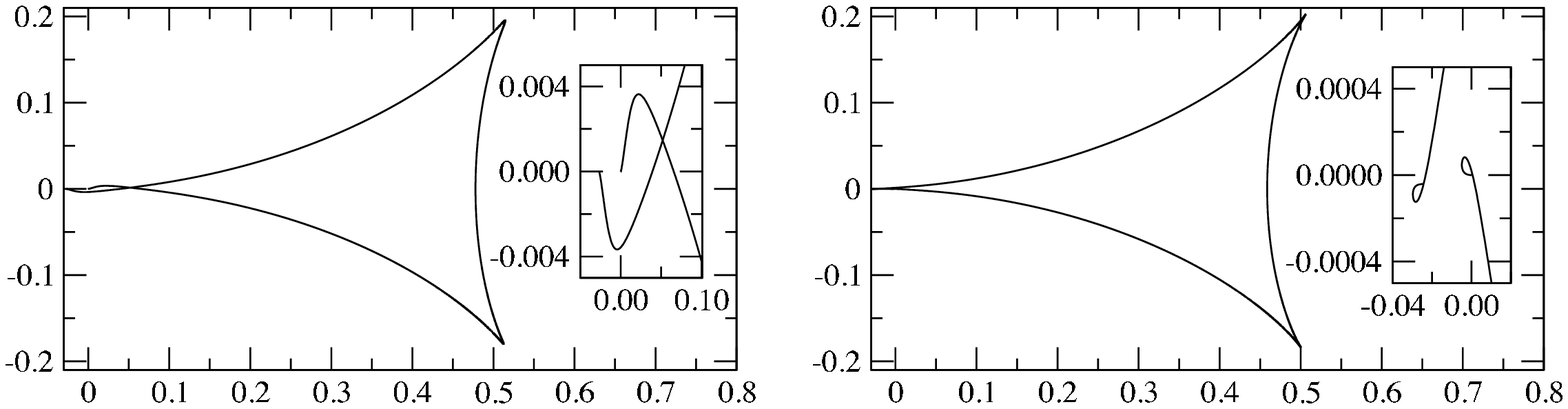}\end{center}
\caption{Tracer trajectory that starts at the origin produced by the scattering with a swimmer that moves along the $x$ axis from left to right, above the tracer. The dimensionless impact parameter is $\widetilde{b}=2$, the swimmer efficiency is $\gamma=1$ and the swimmer model is that of two monopoles (Left) and one dipole (Right). The insets show a zoom near the origin to put in evidence the small displacements. In  both cases, the tracer moves left and down.}
\label{fig.trajectories}
\end{figure}

Figure~\ref{fig.trajectories} shows typical trajectories of tracers for both models. They show the loop-shape also described in \citet{Dunkel2010} for a different model. The net dimensionless displacement 
\beq
\widetilde{\Delta}_i = \widetilde{r}_i(+\infty) - \widetilde{r}_i(-\infty) = \gamma^{-1} \int_{-\infty}^{\infty} d\widetilde{t}\,G_i(\widetilde{t}\HAT{n}+\VECADIM{b}- \VECADIM{r}(\widetilde{t}))
\eeq
is small, effect that can be understood when doing a formal expansion of the equation of motion for small tracer displacements
\beqa
\widetilde{\Delta}_i =\gamma^{-1} \int_{-\infty}^{\infty} d\widetilde{t}\,G_i(\widetilde{t}\HAT{n}+\VECADIM{b})
-  \gamma^{-2} \int_{-\infty}^{\infty} d\widetilde{t}_1\,G_{i,j}(\widetilde{t_1}\HAT{n}+
\VECADIM{b}) \int_{-\infty}^{\widetilde{t}_1} d\widetilde{t}_2\, G_j(\widetilde{t}_2\HAT{n}+\VECADIM{b}) + {\cal O}(\gamma^{-3}). \nonumber
\eeqa
It can be directly verified that the first integral in the expansion cancels both in the case of two monopoles or a dipole. Indeed, this integral corresponds to using the velocity field at a fixed position and, as a result of the head and tail pushing in opposite directions, the effect of the head and the tail cancel and the total induced displacement vanishes.

\begin{figure}
\begin{center}\includegraphics[width=.8\columnwidth]{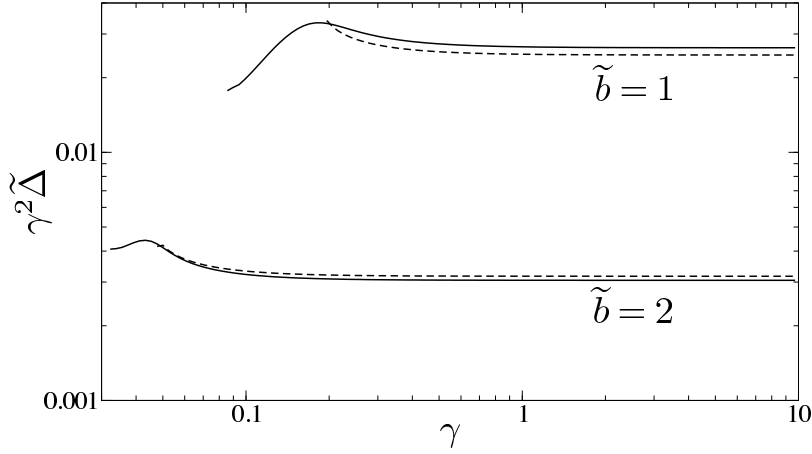}\end{center}
\caption{Net dimensionless displacements $\widetilde{\Delta}$ times $\gamma^2$ as a function of the swimmer efficiency $\gamma$ for two impact parameters: $\widetilde{b}=1$ and $\widetilde{b}=2$. 
Two swimmer models are considerer: two monopoles (solid line) and a dipole (dashed line).}
\label{fig.Deltagamma2}
\end{figure}

Then, the first contribution to net displacement is proportional to $\gamma^{-2}$. Figure \ref{fig.Deltagamma2} shows the net displacement computed by numerical integration of Eq. \reff{ec.motion2} for the two models and two values of $b$ as a function of $\gamma$. The dependence $\Delta\sim\gamma^{-2}$ is obtained  from $\gamma\gtrsim 0.5$ and not only  in the asymptotic regime of large efficiencies. At smaller values of $\gamma$ or small impact parameters, the displacements are large and the tracer approaches the swimmer to the core  of  the singularities, making  predictions not to be trusted. In those cases, more refined swimmer models should be used.
Note that the net displacement does not depend on the sign of $\gamma$ at least in the dominant term. Therefore, there is no difference between pushers and pullers as far as the net displacement of tracers is concerned. We finally remind that the net displacement depends only on $\widetilde{b}$ and $\gamma$ and is independent on the swimmer force or velocity.

Now, considering a tracer in a low concentration suspension of swimmers, the different scattering processes can be studied separately and the total tracer displacement results in the vectorial sum of the sequence of displacements. Let's call $\VEC{\Delta}^{(n)}$ the displacement vector produced in the $n$-th encounter, then the mean square displacement up the $n$-th encounter is
\beq
\langle \Delta R^2 \rangle = \left < \left|\sum_{m\leq n} \VEC{\Delta}^{(m)} \right|^2 \right> .
\eeq
Assuming that  subsequent swimmers move in uncorrelated directions, the cross averages vanish, resulting in 
\beq
\langle \Delta R^2 \rangle = \left< \sum_{m\leq n} \left|\VEC{\Delta}^{(m)} \right|^2 \right>=  a^2 \left< \sum_{m\leq n} \widetilde{\Delta}^2(\widetilde{b}^{(m)},\gamma^{(m)} ) \right> ,
\eeq
where $\widetilde{b}^{(m)}$ and $\gamma^{(m)}$ describe the $m$-th encounter, and the $a^2$ factor appears when substituting $\Delta=a\widetilde{\Delta}$.

To compute the averages, we assume that swimmers are described by a distribution function $f$ such that $f(\HAT{n}, F,\gamma)\, d^2\HAT{n}\, dF\, d\gamma$ is the number of swimmers per unit volume that swim along the direction $\HAT{n}$, with a force $F$, and an efficiency $\gamma$. We assume that, in general, the swimmers in the suspension can have different efficiencies and can swim also at different velocities.
The tracer induced diffusion coefficient equals one sixth the time derivative of the mean square displacement, averaged over the incoming flux of swimmers $V f(\HAT{n}, F,\gamma)$ and the impact parameters
\beqa
\nonumber
D&=& \frac{1}{6} \frac{d}{dt} \langle \Delta R^2\rangle = \frac{1}{6} \int \Delta^2(b,\gamma) f(\HAT{n}, F,\gamma) V\, b\, db\, d\theta\, d^2\HAT{n}\, dF\, d\gamma\\
\label{DInt}
&=& \frac{a^4}{6} \int \widetilde{\Delta}^2(\widetilde{b},\gamma) f(\HAT{n}, F,\gamma) V\, \widetilde{b}\, d\widetilde{b}\, d\theta\, d^2\HAT{n}\, dF\, d\gamma. \label{Dgammadistribution}
\eeqa

If all swimmers have the same efficiency $\gamma_0$ the previous expression simplifies to 
\beq \label{eq.D3D}
D= n \langle V\rangle \beta,
\eeq
with
\beq \label{eq.betaInt}
\beta=\frac{a^4}{6} \int \widetilde{\Delta}^2(\widetilde{b},\gamma_0) \, \widetilde{b}\, d\widetilde{b}\, d\theta,
\eeq
where $n=\int f(\HAT{n}, F,\gamma) \,d^2\HAT{n}\, dF\, d\gamma$ is the swimmer concentration and $\langle V\rangle$ is the average velocity in the swimmer sample.  The coefficient $\beta$ can be written as
\beq
\beta = \langle \Delta^2\rangle \sigma /6,
\eeq
proportional to the product of the cross section $\sigma$ and the average square displacement over this cross section, which has been shown to be  typically small. Therefore $\beta^{1/4}$ should not be understood as a single length scale. 

As has been noted above, the net displacement is roughly proportional to $\gamma^{-2}$ and therefore the diffusion coefficient is proportional to $\gamma^{-4}$. More inefficient swimmers produce larger displacements and, therefore, larger diffusion coefficients. The rationale behind this is that at low Reynolds number, the flow field and the induced displacement are proportional to the swimmer force. On the other hand, the net displacement is proportional to the total interaction time, that is inversely proportional to the swimmer speed. Therefore, a slower swimmer, exerting the same force will induce larger displacements because the interaction time increases.

To compute numerically the coefficient $\beta$ a cutoff distance, $b_{\rm min}$ should be included in the impact parameter: physically, swimmers have a transverse radius that impose a minimal impact parameter and, in the models \reff{Gmonopoles} and \reff{Gdipoles}, a minimal distance should be included to not approach the singularity core. Then, $\Delta(b)$ is computed by numerically integrating the differential equation \reff{ec.motion2} and  the coefficient $\beta$ is then obtained from Eq. \reff{eq.betaInt}. As the displacements are larger for smaller impact parameter, the final value of $\beta$ depends singnificantly on $b_{\rm min}$ and a sensible election should be made. 

In the case of {\em E. coli}, $\gamma=0.89$, and the tracers do not reach the singularity cores for impact parameters larger than $\widetilde{b}=0.75$ in the case of the swimmer modelled as two monopoles and $\widetilde{b}=1.0$ in the case of the dipole model. These values, which are close to the bacterial radius, are chosen as the cutoff impact parameters. The resulting coefficients are $\beta^{\rm dipole}/a^4=0.030$ ($\beta=0.48\, \mu{\rm m})^4$ and $\beta^{\rm monopoles}/a^4=0.088$ ($\beta^{\rm monopoles}=1.40\, \mu{\rm m})^4$, where $a=2\,\mu$m has been used.

As it is discussed in Sec. \ref{sec.experiments}, under our experimental methodology is not possible to extract a value for $\beta$ in the bulk to compare with the theoretical predictions. On the other hand, analysis of the experimental results on {\em Chlamydomonas reinhardtii}  by \citet{Leptos2009} allows to estimate $\beta^{\rm C.r.\, bulk}=(5.5\pm0.7\, \mu{\rm m})^4$. This value, however, cannot be compared with the predictions of the simple swimmer models described here, that do not consider  the time dependence of the flow produced by this alga and the higher multipole moments needed to accurately describe the near flow \citep{Drescher2010, Guasto2010}.

\section{Effect of a solid boundary} \label{sec.difusion.boundary}

Near a solid boundary, swimmers decrease their swimming efficiency due to hydrodynamic hindrance. Lubrication forces and far-field images modify the drag coefficient parallel to the wall such that the swimming speed is
\beq \label{def.alpha}
V_B = \alpha(z_B) V_{B0},
\eeq
where $V_{B0}$ is the swimmer velocity in the bulk, $z_B$ is the distance of the swimmer to the surface and $\alpha$ is the parallel drag correction factor, that vanishes when the swimmer gets in contact with the surface and goes asymptotically to 1 at large distances \citep{Brenner1961, Goldman1967, Holmqvist2006, Huang2007}. 
According to Eq. \reff{eq.gamma}, the swimmer efficiency is modified by the drag with the wall as $\gamma(z_B)=\alpha(z_B) \gamma_0$, where $\gamma_0$ is the efficiency in the bulk. Therefore, close to a solid surface,  the diffusion coefficient should increase by a factor $\alpha^{-4}$.

There are, however, other effects that should be considered close to a surface that can either increase or decrease the diffusion coefficient. First, it is known that {\em E.coli} and other pusher swimmers are attracted to and trapped by solid surfaces, and swim parallel  to them for long time \citep{Ramia1993,Frymier1995,Lauga2006,Berke2008,Li2008,Dunstan2012}. Therefore, we consider trajectories that are parallel to the surface at a fixed height $z_B$ and with planar orientation $\HAT{n}$. Also, the experimental setups in Hele-Shaw geometries allow to measure the in-plane diffusion coefficient and, therefore, only the parallel components of $\VECS{\Delta}$ should be considered.

The second aspect to be taken into account is geometrical: close to a surface, the cross section is reduced as swimmers must be above the surface. In the limiting case of a tracer in contact with the surface, the cross section is reduced by one half. We also take into account that there is a  finite  transverse swimmer radius $r_{0B}$, such that if $z_T$ is the initial height of the tracer, the swimmer height is $z_B=z_T+b \cos\theta$ (see Fig. \ref{fig.paramsscatt}). The condition $z_B>r_{0B}$ limits the integration region in the $b$-$\theta$ space in Eq. \reff{DInt}. 

Finally, the presence of the planar surface modifies the flow profile induced by the force distribution. Using the image method, \citet{Blake74} obtained the solution for a force monopole, taking into account the non-slip boundary condition, therefore modifying the bulk solution \reff{eq.Oseensolution}. First, the flow field is smaller because momentum is dissipated by the surface and, second, the flow field is non-isotropic anymore and depends explicitly on the distance of the source to the solid surface.
The flow produced by a force $\VEC{F}$ evaluated at $\VEC{r}_1=(x_1,y_1,z_1)$, is
\beqa
 u_i (\VEC{x}) &=&\frac{1}{8 \pi \eta} \left[ \left( \frac{\delta_{ij}}{r}+\frac{r_ir_j}{r^3}\right)-\left(\frac{\delta_{ij}}{\bar r}+\frac{\bar r_i\bar r_j}{\bar r^3}\right) \right. \nonumber\\
&&\left. +2h(\delta_{jx}\delta_{kx}+\delta_{jy}\delta_{ky}-\delta_{jz}\delta_{kz} )\frac{\partial}{\partial \bar r_k}\left\{ \frac{h \bar r_i}{\bar r^3}-\left(\frac{\delta_{iz}}{\bar r}+\frac{\bar r_i \bar r_z}{\bar r^3} \right)\right\}\right] F_j ,
\label{Eq.StokesletPared}
\eeqa
where $\VEC{r}_2=\{x_2,y_2,h\}$ is the position of the force point, $\VEC{r}=\{x_1-x_2, y_1-y_2,z_1-h\}$ is the vector from the singularity to the observation point, and $\VEC{\bar r}=\{x_1-x_2,y_1-y_2,z_1+h\}$ is the vector from the position of the image to the observation point.
In the coordinate axes we are using, $x$ and $y$ are the planar directions and $z$ is the perpendicular direction to the plane. This expression, simplified after using that the force is parallel to the surface, allows to obtain the tracer equation for the dipole and two-monopole models.

\begin{figure}
\begin{center}\includegraphics[width=.8\columnwidth]{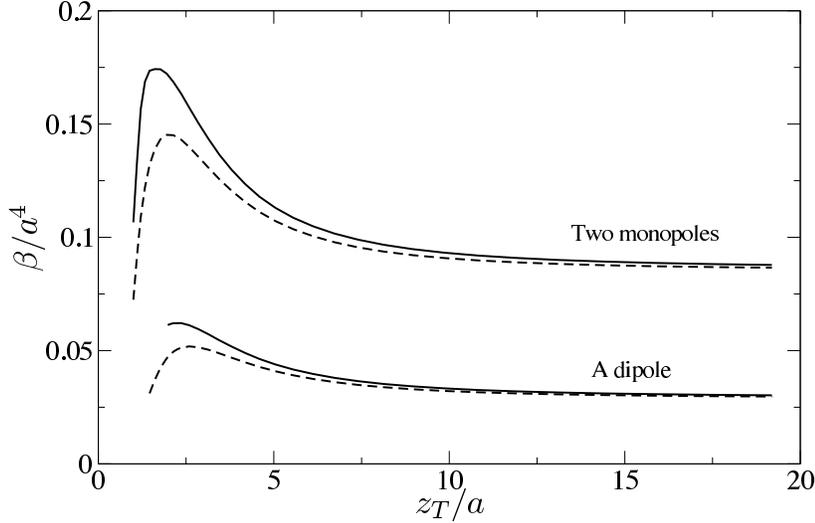}\end{center}
\caption{Effect of the tracer height, $z_T$, on the $\beta$ value. Two swimmer models are considered: two monopoles and a dipole. The calculations are done considering the enhanced drag coefficient (solid line) or without it (dashed line).
The numerical parameters are those that model the {\em E. coli}, that is, $a=2\,\mu{m}$, $\gamma=0.89$, $r_{0B}/a=0.25$, and the cutoff impact parameter is $b_{\rm min}/a=0.75$ for the two-monopole model and $b_{\rm min}/a=1.0$ for the dipole model .
}
\label{fig.BetaWall}
\end{figure}

To study the consequence of these additional effects due to the surface, the $\beta$ coefficient will be first computed without considering the modification of the swimmer efficiency close to the surface. Again, the dominant contribution to the net displacement goes as $\gamma^{-2}$, but the numerical integration of the tracer equation of motion gives two new features. First, if the tracer is close to the surface, the non-slip boundary condition included in Eq.~\reff{Eq.StokesletPared} implies small velocities and therefore small displacements. But, the presence of the surface breaks the symmetry of the flow and, as a result, the trajectories depart more from closed loops as the back-and-forth motion is not symmetric. As a result, the net displacement can increase at a finite distance from the wall and then decreases to approach the bulk values. Integrating over the allowed impact parameters and  azimuthal angles the $\beta$ coefficient is obtained, which  shows a maximum for a finite tracer height (see Figure \ref{fig.BetaWall}). This unexpected result, obtained even in absence of the enhanced displacement due to the swimmer inefficiency near walls shows that the perturbation of the loop-like trajectories compensates the other two effects near walls that tend to decrease the tracer displacement.

Finally, when the effect of the enhanced drag coefficient is included, the $\beta$ coefficient also shows, with larger numerical values, first an increase of $\beta$ with height, reaching a maximum value for a finite height, to further decrease for larger distances, approaching the bulk value. The results for both swimmer models are presented in Fig. \ref{fig.BetaWall} as a function of the tracer height $z_T$. 

Considering the previously described numerical values to model the {\em E. coli} plus the bacterial radius of $r_{0B}=0.5\,\mu {\rm m}$, the maximal diffusivity is obtained when the tracer is located at a height $z_T\sim4.7\, \mu{\rm m}$. At this height $\beta$ is increased by a factor 2 compared to the bulk value. That is, the induced diffusivity is doubled close to the surface. 
Both swimmer models, although producing qualitatively similar results, predict different values for $\beta$. In this respect, the two-monopole model [described by the singularity distribution in Eq. \reff{Gmonopoles}] is more precise than the dipolar model as it  gives a more detailed description of the near field (actually, as it was shown by \citet{Drescher2010} even three monopoles were necessary to describe in detail the near field produced by some microorganisms). Also, the two-monopole model allowed to use  smaller cutoff values, implying larger displacements. In consequence, in what follows, we will concentrate only on the two-monopole model.

\subsection{Confinement by two solid walls} \label{sec.difusion.confinement}

Now we study the effect on induced diffusion in a confined geometry i.e. when the bacteria and the passive tracer are placed between two solid walls. This will correspond to a situation studied experimentally in the next chapter. The theoretical calculations are similar to the previous one. However they are slightly more involved because the image system for point forces between two parallel walls is complex, as the images reflect at the walls an infinite number of times, increasing the complexity of the image system at each reflection \citep{Blawzdziewicz2002}. As a first approximation, we make only one reflection at each wall. That is, the image system presented in Eq. \reff{Eq.StokesletPared} is repeated for the top wall as well. The reflection on the top wall breaks the back-and-forth symmetry even more resulting in an increase of the net displacements compared to the case with just one surface. 

\begin{figure}
\begin{center}\includegraphics[width=.8\columnwidth]{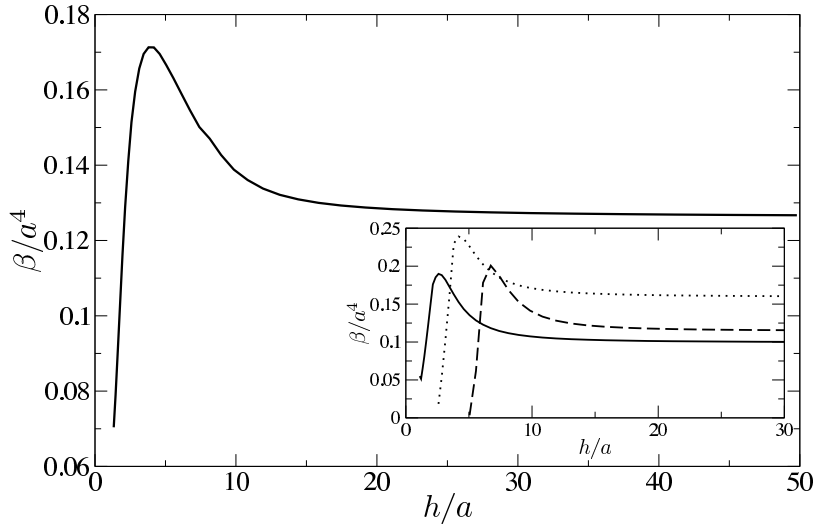}\end{center}
\caption{Effect of the confining height of the system $h$ on the $\beta$ value. The computations are done with the two-monopole model. The main figure shows the predicted values for a buoyant tracer averaged over the Boltzmann distribution of  tracer heights. The inset shows the predicted values for fixed initial tracer heights: $z_T/a=1$ (solid line), $z_T/a=2.5$ (dotted line), and $z_T/a=5$ (dashed line).
The numerical parameters are those that model the {\em E. coli}, that is, $a=2\,\mu{m}$, $\gamma=0.89$, $r_{0B}/a=0.25$, and the cutoff impact parameter $b_{\rm min}/a=0.75$. To compute the Boltzmann averages, the mean tracer height is that of experiments $(k_BT/m^*g)/a = 1.85$.}
\label{fig.BetaTwoWalls}
\end{figure}

The inset of Figure \ref{fig.BetaTwoWalls} shows the effect of the confinement height $h$ on the  $\beta$ coefficient for different fixed tracer heights $z_T$. The main figure shows the effect of confinement averaged over the Boltzmann distribution of tracer heights with numerical values given by the experiments (details are given in Sec. \ref{sec.experiments}). 
First, at short confinement heights, the induced diffusivity increases, reaching a maximum and later it decreases to reach the value of an unconfined system, subject only to the effect of the bottom surface. Both the maximum diffusivity and the height at which it is reached depend on the tracer height. In summary, it is predicted then that, close but not in contact to the solid surface, the induced diffusivity is enhanced under confinement in Hele-Shaw cells.

\section{Experiments on enhanced diffusion}\label{sec.experiments}
To test the theoretical predictions made previously, we performed series of experiments on passive tracers activated by swimming bacteria.
The goal here is to test the validity of Eq. (\ref{eq:Diffusionatthewall}), varying the bacterial density and the mean velocity over a significant range.
A crucial point in the previous report by \citet{Mino2011}  was that a distinction has to be made between active swimmers performing ballistic trajectories and the non-active ones, displaying diffusive motion. Under these conditions,  the effect of enhancement in the diffusion of passive tracers can be assessed quantitatively. In the bulk, because of the 3D motion, bacteria rapidly go out of focus, and there is not enough data to make the separation between the two populations. Consequently, we have restricted ourselves to measure close to the surfaces where the separation between active and non-active bacteria can be made without ambiguity. Also, in the previous experiments, the dependence of the swimmer velocity on the diffusion coefficient was only investigated by taking bacteria at different stages of their life cycle i.e. the post division fast swimmers (1N cells)  and the pre-division slow swimmers (2N cells). In this way, a factor 2 between the corresponding mean velocities was obtained. However, 1N and 2N populations have different shapes which from the previous theoretical analysis, can be a questionable issue since a change in shape would imply a change in the efficiency parameter $\gamma$. If there were a variability in $\gamma$ then Eq. (\ref{eq:Diffusionatthewall}) is no longer valid and the diffusion coefficient is given instead by the more involved expression \reff{Dgammadistribution}.
Finally, the advantage of the present experiment is that by varying the swimming velocity, relation (\ref{eq:Diffusionatthewall}) can be studied more extensively. 

Another question addressed in the theoretical sections is the influence of confinement, i.e. the presence of another surface  limiting  the height of the chamber. From the experiments it is possible to extract the effect of confinement in the Brownian diffusivity as well as in the enhanced diffusion due to the presence of bacteria.

The bacteria used here are  Wild-type \textit{E. coli} (ATCC 11105). They were grown and prepared following the protocol described in Appendix \ref{appA}.

Observation on the bacterial suspension was performed using a \emph{Z1} inverted microscope from \emph{Zeiss-Observer}. Images and videos were captured using a \emph{Pixelink PL-A741-E} CCD digital camera, connected to and controlled by a computer which stores the images that will be post-processed. The CCD chip has a maximum resolution of $1024\times1280$ pixels$^2$ and can run at 10 frames$/$s full frame. To gain speed of acquisition, we reduced the visualisation field to $600\times800$ with a rate of 20 frames$/$s. 

For experiments on induced diffusion, the principle is to follow the motion of passive particles in a bacterial suspension.  A sketch of the chamber is shown in figure \ref{fig:Gen_Des} (a). A 10 $\mu$l drop of the suspension containing the W wild-type \emph{E. coli} bacteria \citep[see][]{Archer2011} and the tracer latex beads of diameter $d = 2 \mu m$, were placed in a transparent chamber on the visualisation stage of the microscope.  The chamber is made of two cover-slips separated by a typical distance $h$,  ranging between $5$ and $ 110 \mu m$ (see  Appendix \ref{appA} for details).

\begin{figure}
\centerline{\includegraphics[width=0.7\columnwidth]{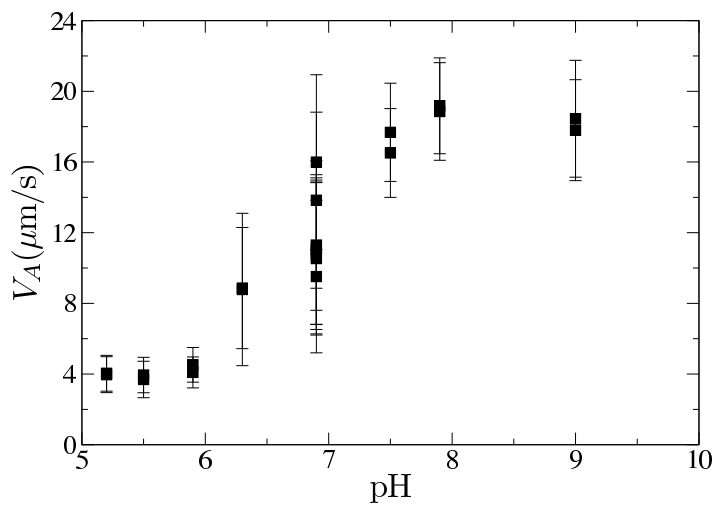}}
  \caption{ Effect of the pH in mean velocity $V_A$ of the active population using MMA solution. Error bars represent standard deviations.}
\label{fig:VA_pH}
\end{figure}

\begin{figure}
  \centerline{\includegraphics[width=1\textwidth]{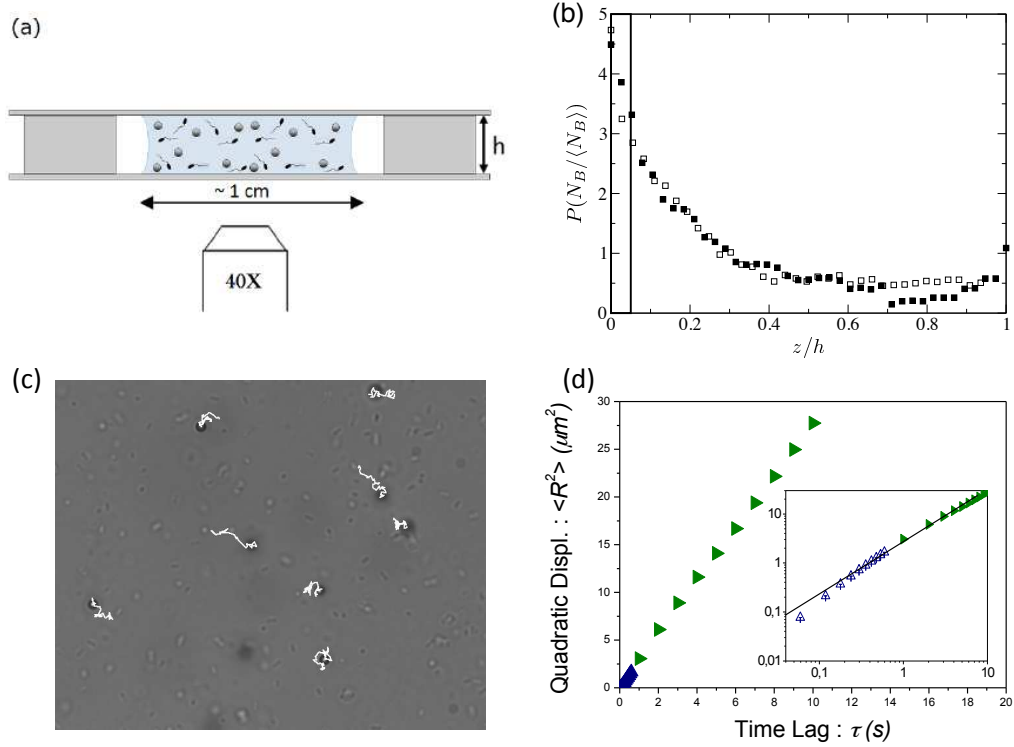}}
  \caption{ (\textit{a}) Sketch showing a  a lateral view of the chamber observed on the microscope. For these experiments a 40X objective (Aperture number AN 0.65) was used, given a visualisation field of $96\times128\,\mu$m$^2$. Observation are performed in the centre of the drop, far away from the drop border. (\textit{b}) Distribution of bacteria as function of height in buoyancy solution (Minimal Motility Medium at pH 7) for two bacterial concentration $n$: $4.5\times10^8$bact/ml (solid squares) and $7\times10^8$bact/ml (empty squares), corresponding to optical densities OD 0.7 and 1, respectively. The cell height is $h=110\,\mu{\rm m}$.
(c)  picture of tracer trajectories.
(d) quadratic displacement as a function of time lag at two temporal resolutions.
}
\label{fig:Gen_Des}
\end{figure}

As discussed previously our goal is to produce bacterial samples at different concentrations and with different mean swimming velocities. In the work presented by \citet{Minamino2003} the effect of pH  in the motility of \emph{E.  coli} and \emph{Samonella} motility was studied. They varied the pH between 5 and 7.8, showing that in the presence of potassium acetate, there is a significant variation of the swimming speed reaching a maximum at pH 7. Inspired by this result, we used a similar protocol  to vary the velocity of the bacteria (see Appendix \ref{appA}).  We obtained a mean velocity variation   from 4 $\mu$m/s at low pH up to 20 $\mu$m/s at high pH, that is, an increase of the mean velocity by a factor 5 (see Fig. \ref{fig:VA_pH}). However it is important to notice that the results presented in \citet{Mino2011} were obtained in non-buoyant condition using a 1:1 mixture of Minimal Motility Medium (MMA) and Percoll, a nanoparticles suspension \citep[]{Laurent1980a}. We actually noticed that the viscosity of the MMA-Percoll mixture changes strongly with pH and forms a gel at low pH, therefore, we could not obtain a density matching between the solution and both the beads and the bacteria while keeping the viscosity fixed.
Consequently, the experiments presented here were done with MMA only (without Percoll) and we tested by direct rheometric measurements that in all cases the fluid stays Newtonian. The pH variation does not affect the viscosity of the fluid with a value, at 25$^\circ$C, close to the one of pure water (see Appendix \ref{appA}).
 In this context, the latex particles (density 1.03 g/cm$^3$) will sediment and stay preferentially close to the bottom wall. In figure \ref{fig:Gen_Des}(b),  bacterial concentration profiles are shown in a situation where the chamber height is h = 110 $\mu$m in buoyant conditions. We observe an increase of concentration close to the bottom surface and the corresponding visualisation region is displayed in a rectangle on the same figure. Also an increase of the bacteria concentration near the top wall can be noticed, phenomenon that has been reported previously by \citet{Frymier1995,Lauga2006,Berke2008,Li2008}, as a consequence of the trapping of bacteria at the surface by hydrodynamic interactions

Given the specific protocol explained in Appendix A, once the chamber is placed under the microscope, the steady distribution of bacteria in the vertical direction is reached very rapidly. The experiments start few minutes  later  in order  to let the latex beads sediment. Figure \ref{fig:Gen_Des}(c) shows a snapshot of several latex beads among the bacteria close to the bottom surface and we superpose the corresponding  tracks for 30 s.  In Fig. \ref{fig:Gen_Des}(d) is shown an averaged quadratic displacement as a function of the time lag $\tau$ that the tracers are tracked, obtained at two different frame-rates of acquisition. This indicates that after 0.5~s , a  diffusive motion is reached. The diffusion coefficients extracted by linear fit of the quadratic displacement versus time lag are obtained by analysing $300$ images, taken at 1 frame per second.  The time lags used for the fit are between 1-10 s.  For all the data,  the bacterial motion is analysed using  20 s videos at 20 images per second.  The active bacteria concentration $n_A$ is determined by characterising  the tracks with the method described in  \citet{Mino2011}, using as indicators the mean persistence angle and the exploration radius. For each experiment, the mean bacteria velocity $V_A$ is computed from the mean track length over the total visualisation time.

\subsection{Diffusivity enhancement close to the bottom surface }
\begin{figure}
  \centerline{\includegraphics[width=0.7\textwidth]{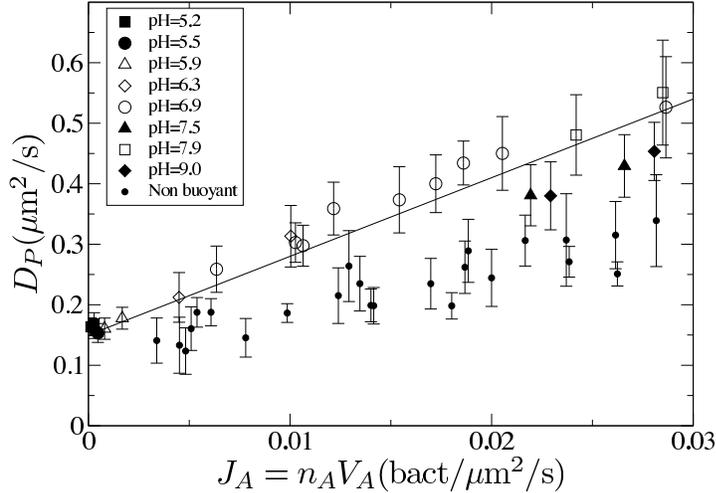}}
  \caption{Enhanced diffusivity $D_P$ of passive tracers as a function of $J_A$ in buoyancy conditions. Squares represent tracer of $2\,\mu$m diameters. Each symbol represents an experiment performed with different pH. Black circles represent the result obtained under isodense (non-buoyant) condition for passive tracers of $2\,\mu$m diameter \citep[see][]{Mino2011}.}
\label{fig:Enhance_Bact_MMA}
\end{figure}

In this first series of experiments the chamber height is maintained at its higher value of 110 $\mu$m. As described in detail in Appendix A, the change of pH changes the mean bacterial velocity $V_A$. Systematic measurements of the diffusion coefficient of tracers were made for suspensions prepared at pH $5.2$, $5.5$, $5.9$, $6.3$, $6.9$, $7.5$, $7.9$ and $9$. For each pH, the bacterial concentration was varied by preparing suspensions at different OD (Optical density) ranging from $OD=0$ up to $OD = 1$. As discussed before, for each experiment, the corresponding values of the particle diffusivity $D_P$, the mean active bacterial velocity $ V_A$ and the mean active density $n_A$ were measured. Figure \ref{fig:Enhance_Bact_MMA} shows the relation between passive tracer diffusion $D_P$ and the active flux $J_A= n_A V_A$.  Different symbols represent different pH, except for the solid circles which correspond to the results presented by \cite{Mino2011}.  The data collapse on a straight line as it is proposed in Eq. (\ref{eq:Diffusionatthewall}) where the Brownian diffusivity near solid boundaries is
\beq
D_P^B=\alpha D_B=\frac{\alpha k_B T}{3\pi \eta d}.
\end{equation}
Here, $k_B$ is the Boltzmann constant, $T$ is the absolute temperature, $\eta$ the viscosity of the liquid, $d$ the diameter of the particle and $\alpha<1$ is the parallel drag correction factor introduced in \reff{def.alpha}.
From the data presented in Fig. \ref{fig:Enhance_Bact_MMA}, the fitted value for $\alpha$ is $0.80\pm0.08$ and $\beta$ is $13.0 \pm0.7\,\mu$m$^4$, where the last one gives the influence of bacterial motion in the tracers diffusion. 

Now, comparing with the values reported previously in a non-buoyant condition \citep{Mino2011} the value $\alpha$ found for passive tracer of $2\,\mu$m was $0.74\pm0.03$, and for $\beta$ was $5.0\pm0.4\,\mu$m$^4$.  It can be noticed that for small heights $\beta$ in a buoyant condition it is significantly higher than the one found for non-buoyant conditions. This result can be observed in Fig. \ref{fig:Enhance_Bact_MMA}, where the previous results in non-buoyant conditions (black circles) also are plotted.

To make a numerical comparison with the theoretical model, we must note that although the tracers are buoyant, they are not in contact with the surface due to thermal Brownian motion. In fact, the tracer heights are distributed according to the Boltzmann distribution $\rho\sim \exp(-m^*g z_T/k_BT)$, where $m^*$ is the buoyant mass, with an average tracer height  $k_BT/m^*g=3.7\,\mu{\rm m}$. Then, the $\beta$ coefficients shown in Fig. \ref{fig.BetaWall} must be averaged over the tracers distribution, resulting in $\beta^{\rm avg} = 2.0\,\mu{\rm m}^4$ when the two-monopole model is used. The predicted numerical values are  smaller than the observations. The difference could be attributed to the simplified swimmer models, which do not treat adequately the near field flows. 

\subsection{Variation of active diffusivity with the confinement}

\begin{figure}
\centerline{\includegraphics[width=0.9\textwidth]{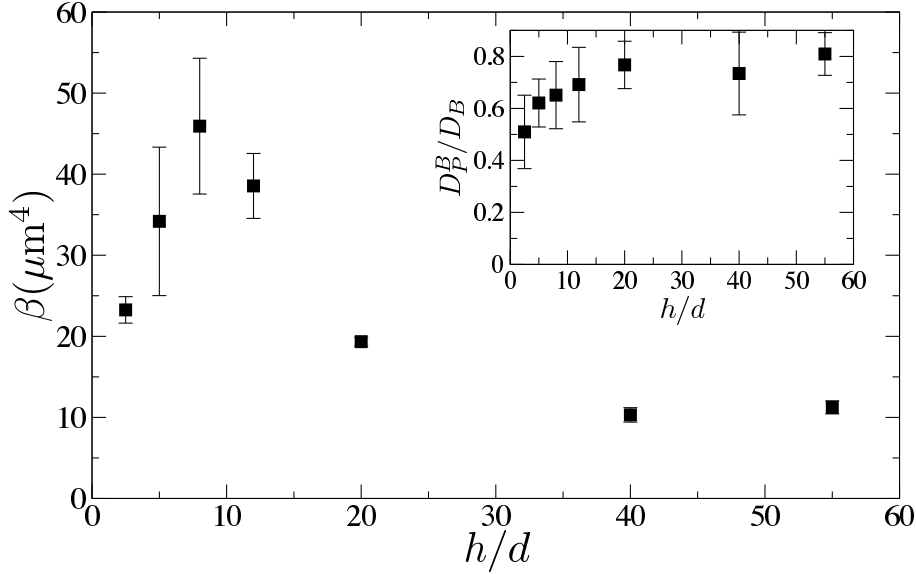}}
  \caption{
  Effect of the confinement on the $\beta$ value. Experiment was performed in MMA at pH 7, using $2\,\mu$m diameter passive particles. 
  Inset: Effect of the confinement in the value $\alpha$  for $2\,\mu$m diameter passive tracers. $D^B_{P}$ represent the diffusion of particles without bacteria and $D_B$ the thermal diffusion in the bulk.}
\label{fig:Beta_diff_h}
\end{figure}

Now we explore the effect of varying the confinement height $h$ between the top and the bottom walls. In these experiments, the pH value is maintained at $pH = 6.9$ and the OD is varied from 0 to 0.9. 
At each value of $h$, we extract by fitting linearly the relation between $D_P$ and $J_A$, the Brownian diffusivity coefficient in the absence of bacteria  ($D_P^B$)  and the slope $\beta$ representing the activation effect. The results for various confinement heights  $h$ are presented in Fig. \ref{fig:Beta_diff_h}. In the Inset, we plot the relative Brownian diffusivity $D_{P}^B/D_B$, where $D_B$ is the theoretical thermal diffusion in the bulk. It can be noted that  the value of $D_{P}^B/D_B$ decreases when the separation between the two wall decreases, which is in good agreement with the experimental results presented by  \citet{Lobry1996}, in which they measured the transverse Brownian motion of passive particles in a confined geometry.
In the main figure we display the effect of confinement in the enhanced diffusion, given by the slope $\beta$ in Eq. \reff{eq:Diffusionatthewall}. $\beta$ increases significantly when the upper wall is near to the bottom one, presenting a maximum at height $h\sim 8\,\mu$m. This maximum is approximately four times the value for unconfined system.

The observed enhancement in the diffusion under confinement is similar to the predicted effect in Sec. \ref{sec.difusion.confinement}. To make a quantitative comparison, the $\beta$ coefficients showed in the inset of Fig. \ref{fig.BetaTwoWalls} must be averaged over the tracer distribution, results that are displayed in Fig. \ref{fig.BetaTwoWalls}.
There is a remarkable similarity among the two curves. When decreasing the confining height, both show a maximum in $\beta$ at a finite height  to further decrease when the two plates are very close. The position of the maximum for the model, $h\sim 11\,\mu{\rm m}$, is comparable with the experimental value.  However, the predicted values and the observed enhancement are smaller than the experimental values.
This quantitative discrepancy can be attributed to the simplicity of the swimmer models,  the non-uniformities of the bacterial concentration near the surface and the approximation made to calculate the image system between two surfaces (see Sec. \ref{sec.difusion.confinement}).

\section{Conclusion and Discussion}\label{sec.conclusions}

Experimentally, in a mixture of swimmers and tracers, it has been observed that at low swimmer concentration, the agitation created by their movement  produces an enhanced diffusive motion on the tracers, with a diffusion coefficient that grows linearly with the swimmer concentration. A kinetic theory explanation, that uses the properties of the hydrodynamic interactions at low Reynolds number, is given to this experimental observation. The successive displacements produced by uncorrelated swimmers produce a diffusive motion, with a diffusion coefficient that is proportional to the swimmer concentration, their mean velocity and a constant $\beta$ that scales as the scattering cross section and the mean displacement squared in individual swimmer-tracer encounters. A dimensional analysis shows that the displacements grow for decreasing swimmer efficiencies, that is measured as the swimmer velocity times its size and the fluid viscosity, divided by the force of the swimmer. It results, then, that inefficient swimmers produce large tracer diffusion. The calculations are done for simple swimmer models described as a force dipole or two monopoles. More complex models can be used as well, leading to the same qualitative conclusions on the scaling of the diffusion coefficient with the bacterial concentration and mean velocity, although with different expressions for $\beta$. A separate analysis, though, should be used in the case of swimmers that have time dependent strokes as is the case of {\em Chlamydomonas reinhardtii}.

The swimming efficiency is decreased near solid surfaces, where the modification of the viscous drag produces smaller swimming speeds. Near surfaces there are, however, other effects that lead to either an increase or decrease of the induced diffusion. Numerical calculations, considering all these effects, show that the induced diffusion is enhanced near solid surfaces and present a maximum at a finite distance form the wall.

To verify these results, experiments with {\em Escherichia coli} were performed in confined Hele-Shaw geometries. Changing the solution pH it was possible to modify substantially the swimming speed, and together with a variation in the bacterial concentration, it was possible to deduce the $\beta$ coefficient from the measurements of the diffusion coefficient. Experiments were done under buoyant conditions and therefore surface effects are present. To test the predicted enhancement of tracer diffusion near surfaces, the confinement height was changed, finding that $\beta$ increases by 4 times when the cell height is only $8\,\mu{\rm m}$. There is a good qualitative agreement with the theoretical calculations, that predict, however, smaller increases.

It must be remarked that the swimmer models used to obtain the numerical predictions of $\beta$ were very crude (a dipole of forces or two force monopoles) which are expected to describe correctly the fluid motion only at large distances. It is then remarkable that the numerical prediction give the correct order of magnitudes and the mean features produced by the solid surfaces. To obtain quantitative predictions, the same theoretical framework can be used but more accurate swimmer models should be considered, including higher multipole moments, to compute the tracer displacements.


We thank A. Honerkamp-Smith for valuable comments. This research is partially supported by Fondecyt 1100100, Anillo ACT 127, and ECOS C11E04 grants. J.D. thanks the support of a CONICYT grant (Becas Chile). G.M. and E.C. acknowledge the financial support of the  the Pierre-Gilles de Gennes Foundation and the Sesame Ile-de-France ``SMAC''. 

\appendix

\section{Bacteria growth and working suspensions}\label{appA}

In this Appendix, we describe how the samples were prepared. We present the characterisation of the minimal Motility Medium (MMA) and the bacteria cultivation as well as the beads and chamber preparation.

\subsection{Minimal Motility Media (MMA)}

The choice of the ambient medium is always a difficult issue as it may strongly modify bacterial behaviour in terms of chemotaxis, ambient fluid viscosity or response to pH. In the literature, several protocols were designed to achieve different goals. For example, for \emph{E. coli}, \citet{Berg1972} defined an specific protocol to study the chemotactic effects of amino-acids in the bacterial response. A different protocol was chosen by \citet{Lowe1987} as they sought to change the viscosity of the suspending medium in order to study the rotation of flagella bundles with \emph{Streptococcus}. Note that in general, the biological conditions impose a very narrow window of parameters and in practice it is always difficult to conduct experiments that change one parameter at a time as soon as a set of working conditions is established. In the present study, we consider a Minimal Motility Medium (MMA) as the swimming medium. The use of MMA has several advantages. First, it contains a sufficient amount of nutritional elements to preserve the bacterial metabolism, but cell division is strongly reduced, thus allowing control over density of the bacterial population and limiting the influence of chemotaxis. Second, this medium enables adjustment of bacterial velocity by modifying the pH of the solution, acting directly on the molecular proton motor that activates the flagella rotation \citep[see][]{Minamino2003}. A third advantage is that for different values of pH, the MMA solution remains Newtonian and it keeps  physical properties similar to water as far as density and viscosity are concerned.

MMA was prepared as followed: stock solutions (1 M NaH$_2$PO$_4$ and 1M K$_2$HPO4) were mixed to generate phosphate buffers at a given pH (from 5.5 to 8). The stock phosphate buffers were diluted to a final concentration of 10 mM in the presence of 0.1 mM K-EDTA, which favours motility \citep{Adler1967, Adler1973} and 20 mM sodium lactate as an energy source. In some cases 1\% glucose was added to prevent the effects of deoxygenation. Potassium acetate at 34 mM was also present in order to obtain bacteria with swimming velocities sensitive to pH \citep{Minamino2003}.

In order to characterise the viscosity of pure MMA solutions at various pH, an Anton Paar Physica MCR Rheometer was used. For each solution, the shear rate was varied from 10 to 100s$^{-1}$, and 20 points were measured at 25$^{\circ}$C. The procedure was repeated three times. The pH variation does not affect the viscosity of the fluid which stays Newtonian with a value, at 25$^{\circ}$C, close to the one of pure water . 

\subsection{Bacteria cultivation}

Wild-type \textit{E. coli} (ATCC 11105) were grown overnight at $25^\circ$C in rich medium LB (Luria broth). The next day, part of this culture was diluted 100 times in LB, allowing logarithmic growth to restart. Several protocols were performed in order to obtain swimming bacteria with maximal velocity.
Some authors claim that the maximum velocity is reached in the mid-exponential phase of the growth curve \citep{Staropoli2000} whereas others, and also our observation, found that new born bacteria from the post exponential phase run faster \citep{Amsler1993, Prub1997}.
Thus once an optical density of 0.6 was reached (measured with spectrometer from Hitachi), the bacterial suspension was washed twice by centrifugation (500 rpm during 10 min) in MMA to eliminate all LB traces, and finally resuspended in MMA solution (pH 7) for at least an hour, allowing division to end, and obtaining a population enriched in 1N bacteria, also called ``baby cells". In some cases, the ``baby cells" were sorted by low speed centrifugation.
The optical density of the resulting suspension was measured in order to prepare, by serial dilutions, samples of bacteria at different concentrations. Finally, when bacterial suspensions were introduced in the microchamber, Polyvinylpyrrolidone (PVP) at 0.005\% was added to the suspension to avoid stickiness.

\subsection{Beads preparation}

To study the effect of bacterial activity on the diffusivity of passive tracers, latex beads of 2 micron diameter (Beckman-Coulter, density  $\rho$=1.027 g/ml) were added to the suspensions. These spherical particles come in an aqueous medium containing surfactant that was eliminated by 3 successive centrifugations in MMA. Finally they were resuspended in MMA containing 0.005\% PVP.

\subsection{Chamber preparation}

The chamber is made of two coverslips separated by a typical distance $h$. The largest distance  $h=110\,\mu$m  is achieved by using two calibrated coverslips N$^{\circ}$0  (0.085 to 0.13 mm thick) as spacers. For smaller confining heights, we use as spacer calibrated beads previously dispersed in the suspension. The action of capillary forces of the outer rim of the droplet  will hold tight the bottom plate and the top coverslip at the right distance from each others. In this way we can achieve confinement heights ranging between $5$ and $ 110 \mu m$. 

Finally, to avoid the stickiness, coverslips were coated with ethanolic PVP (20 mg/ml of PVP in pure ethanol at 20 mg/ml) and dried.

\end{document}